# Lightweight holographic near-eye display system with self-charging capability using solar energy

Changyu Wang, Yuan Xu, Hong Xu & Juan Liu*

**Beijing Engineering Research Center for Mixed Reality and Advanced Display, MoE Key Lab of Photoelectronic Imaging Technology and System, School of Optics and Photonics, Beijing Institute of Technology, Beijing 100081, China**

**juanliu@bit.edu.cn*

## Abstract

Near-eye display plays an important role in emerging spatial computing systems, providing a distinctive visual effect of virtual-real fusion. However, its application for all-day wear is greatly limited by the bulky structure, energy expenditure, and continuous battery heating. Here, we propose a lightweight holographic near-eye display system that takes advantage of solar energy for self-charging. To achieve the collection of solar energy and near-eye display without crosstalk, holographic optical elements (HOE) are used to diffract sunlight and signal light into a common waveguide. Then, small-area solar cells convert the collected solar energy and power the system. Compact power supply components replace heavy batteries, contributing to the lightweight design. The simple acquisition of solar energy provides the system with sustainable self-charging capability. We believe that the lightweight design and continuous energy input solution will significantly promote the popularity of near-eye display in our daily lives.

# 1 Introduction

As a fundamental advanced display technology of emerging spatial computing device, near-eye display offers an immersive or perspective visual effect, demonstrating enormous potential for widespread applications. Nevertheless, the long-term wearing comfort and convenience of near-eye display system influence its further integration into daily work and life of general consumers. On one hand, system weight is an important factor affecting wearing comfort[1,2]. Typically, the near-eye display devices of the current commercial products weigh hundreds of grams, with high-capacity batteries accounting for over half of the weight, resulting in wearing pressure and structure design difficulties. On the other hand, the limited battery life, about few hours, leads to frequent charging, making continuous use of the system troublesome or even impossible. Besides, the continuous heating of battery during operation also negatively affects system performance and user experience, increasing the demand for effective thermal management.

There have been early efforts to address the weight burden and endurance capability of near-eye display system. To reduce the weight and volume of the main system, three effective solutions include utilizing lighter image sources such as micro light-emitting diode[3] and laser beam scanning[4], designing lighter optical combiners including waveguides[5,6], holographic optical elements[7,8], or micro- and nano-optical elements[9,10], as well as separating light sources[11,12] from the main system. Although the above methods can reduce system weight to a certain extent, the decrease in display performance and the inconvenience of separating modules are often unacceptable. Moreover, the optimized system still includes heavy batteries, which greatly affects the demand for lightweight design. In addition, regarding the issue of battery life, researchers propose two solutions: improving energy utilization efficiency through reasonable optical design[7,13] and reducing energy consumption by implementing mobile edge computing[14,15]. Obviously, the above methods can only appropriately extend battery life, so frequent battery charging is still inevitable.

Self-charging eliminates the need for high-capacity batteries and additional charging devices, satisfying the requirement of charging while consuming. Currently, existing self-charging systems typically use biomechanical energy[16-18], thermoelectric energy[19,20], piezoelectric energy[21], and solar energy[22-25]. Solar energy, as a widely available clean energy, is a promising energy source for near-eye display system, but the easy acquisition of solar energy without adding complex structure is also a challenge. Overall, there are some methods to respectively improve the comfort and convenience of near-eye display system. However, there is a lack of fundamental solutions that comprehensively achieve lightweight design and continuous self-acquisition of clean energy from an optical perspective.

In this study, we present a lightweight holographic near-eye display system with self-charging capability by simply capturing solar energy. The utilization of solar capture HOE and waveguide enables the collection of solar energy over a wide range in a compact structure without affecting near-eye display, providing an excellent continuous energy input solution. This not only adds self-charging function to the system, but also avoids the impact of continuous battery heating on user experience and heat dissipation design difficulties. Replacing heavy batteries with small-area solar cells significantly reduces the weight of near-eye display system and promotes the revolutionary evolution from a helmet structure to a wireless glasses structure.

## 2 Results

### 2.1 Working principle

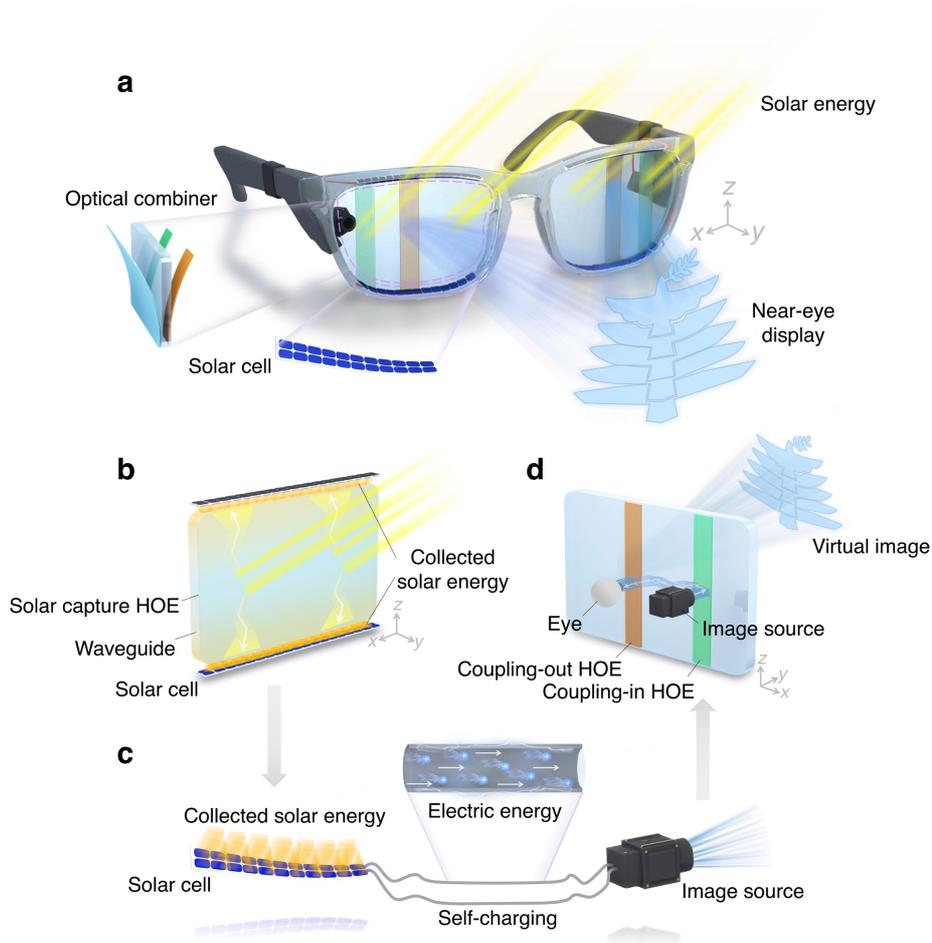

**Fig. 1 | Illustration of the architecture and working principle of proposed system.**
**a**, Architecture of the proposed system. The core component is an optical combiner consisting of a waveguide and three HOEs, one on the outer surface and two on the inner surface. In addition, solar cells are located on both sides of the optical combiner above and below. **b**, Schematic diagram of solar energy collection. The external sunlight is captured by solar capture HOE and enters the optical waveguide, then collected along the z-axis in the form of total internal reflection to the upper and lower sides of waveguide, and finally irradiated onto the solar cells. **c**, Schematic diagram of solar power supply. Solar cells convert collected solar energy into electrical energy and realize self-charging to the image sources. **d**, Schematic diagram of near-eye display. The image source powered by solar energy emits signal light, which is coupled into the waveguide by Coupling-in HOE, then propagates through total internal reflection along the x-axis, and finally is coupled out from the waveguide by Coupling-out HOE, achieving near-eye display.

Figure 1a illustrates the basic structure of the proposed system, which can be divided into two parts: solar power supply module and near-eye display module. The core component is an optical combiner consisting of a waveguide and three HOEs. Regarding one of the two symmetrical eyeglasses, the solar power supply module includes a solar capture HOE (HOE-1), a waveguide, and two solar cells, where HOE-1 is anchored to the outer surface of waveguide. The near-eye display module includes an image source, a coupling-in HOE (HOE-2), a coupling-out HOE (HOE-3), and a waveguide, where HOE-2 and HOE-3 are attached to the inner surface of waveguide. It should be noted that the waveguides of the two modules are the same one and serve as a common carrier for collected solar energy and signal light emitted from the image source. Compared with the traditional holographic near-eye display system, simply adding a thin film and small-area solar cells bring numerous benefits of self-charging capability.

Figures 1b, c, d illustrate solar energy collection, power supply, and near-eye display, respectively. Sunlight and signal light are separately coupled into waveguide by HOE-1 and HOE-2, propagating through different total internal reflection directions along z-axis and x-axis, and finally coupled out at HOE-3 and the edges of waveguide respectively. Solar cells receive collected solar energy and convert it into electrical energy to power the image source. Because of the angle selectivity and wavelength selectivity of volume HOE, only light that meets the Bragg condition will be diffracted with high efficiency, so the optical functions of three HOEs will not interfere with each other.

The collection characteristics of HOE-1 will directly affect the power supply efficiency. On one hand, the secondary diffraction effect will cause part of the collected light to be coupled out in advance and unable to propagate to the edge of the waveguide, thereby reducing the effective collection area. On the other hand, the selectivity of HOE will lead to limitations in collection wavelengths and angles. By partitioning HOE-1 and rigorously designing recording and reconstruction conditions for each partition, the collection angle, efficiency, and effective area of HOE-1 are obviously improved. Different from previous solar energy collectors, we fully utilize the flexibility and freedom of HOE and develop a sustainable real-time solar energy input solution suitable for near-eye display system.

## 2.2 Experimental Results

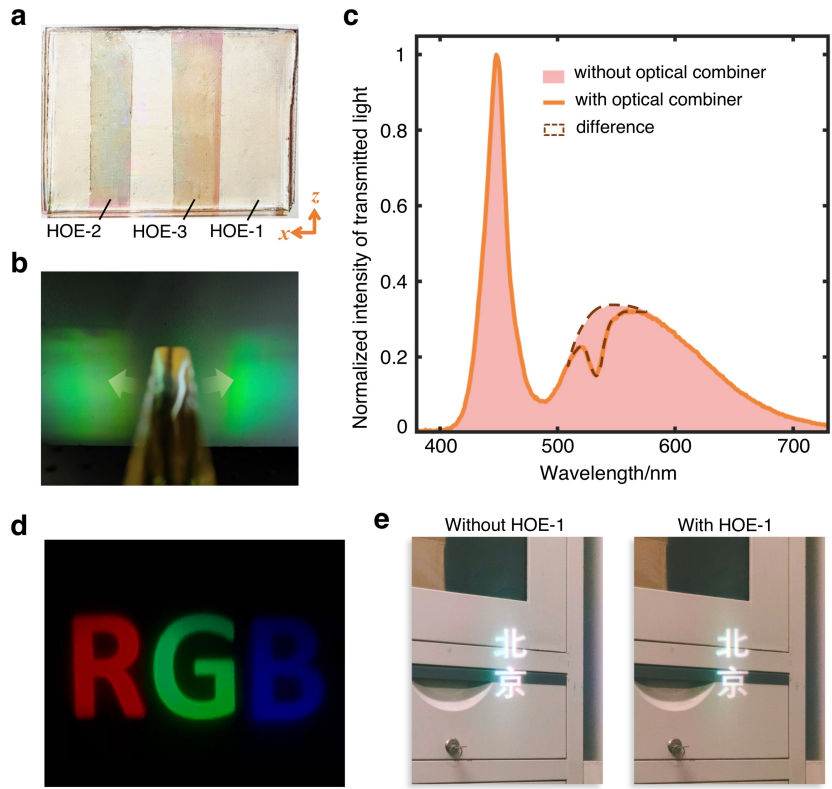

**Fig. 2 | Results of optical combiner and measurements.**
**a**, Prototype of the optical combiner. HOE-1 is located on the front surface, and HOE-2 and HOE-3 are located on the back surface. Since the two eyeglasses are symmetrical, only one of them is used for measurement. **b**, Collected light emitted from the edge of the waveguide. **c**, Comparison of transmitted light spectrum without and with the optical combiner. The pink area and orange solid line respectively represent the transmitted light spectrum without and with the optical combiner. The brown dashed line represents the difference, which is collected by the optical combiner. **d**, Full color near-eye display effect. The letters R, G, and B and their corresponding colors are used to display. **e**, Comparison of near-eye display without and with HOE-1. The AR near-eye display effects of the Chinese characters "Beijing" are shown in the left and right images, respectively, without and with HOE-1.

In order to verify the rationality and feasibility of the proposed system, we have experimentally fabricated a prototype of the optical combiner as shown in Figure 2a. HOEs are fabricated by holographic exposure as designed and attached to a glass substrate, which serves as the waveguide. The fabrication wavelength of HOE-1 is 532nm, and the collected solar energy is designed to propagate in total internal reflection mode along the z-axis inside the waveguide. HOE-2 and HOE-3 are fabricated by synthetic white light formed by 457nm, 532nm, and 639nm lasers, and the signal light is designed to propagate in total internal reflection mode along the x-axis.

On one hand, when using a flashlight to illuminate the optical combiner with white light at the designed incident angle, the collected light emitted from the edge of the waveguide is shown in Figure 2b. Figure 2c illustrates a comparison of the transmitted light spectrum without and with the optical combiner. Define half of the maximum collection efficiency (the ratio of the collected light power to the incident light power) as the threshold for effective collection. At the designed incident angle, the center wavelength of the effective collected light is 532nm and the spectral width is approximately 16nm. Overall, the optical combiner can effectively accomplish the function of solar energy collection.

On the other hand, using a laser projector as the image source, the target image consists of the letters "R", "G", and "B" and their corresponding colors. The full-color near-eye display effect is shown in Figure 2d. Figure 2e shows the impact of HOE-1 on the display effect, when displaying the Chinese characters "Beijing" in the physical environment as augmented reality. The left and right figures show the display effects without and with HOE-1 respectively. It is obvious that HOE-1 causes a slight decrease in the transmittance of the physical environment, but it hardly affects the display effect of the target information. Therefore, the impact of solar energy collection on near-eye display is negligible, as expected.

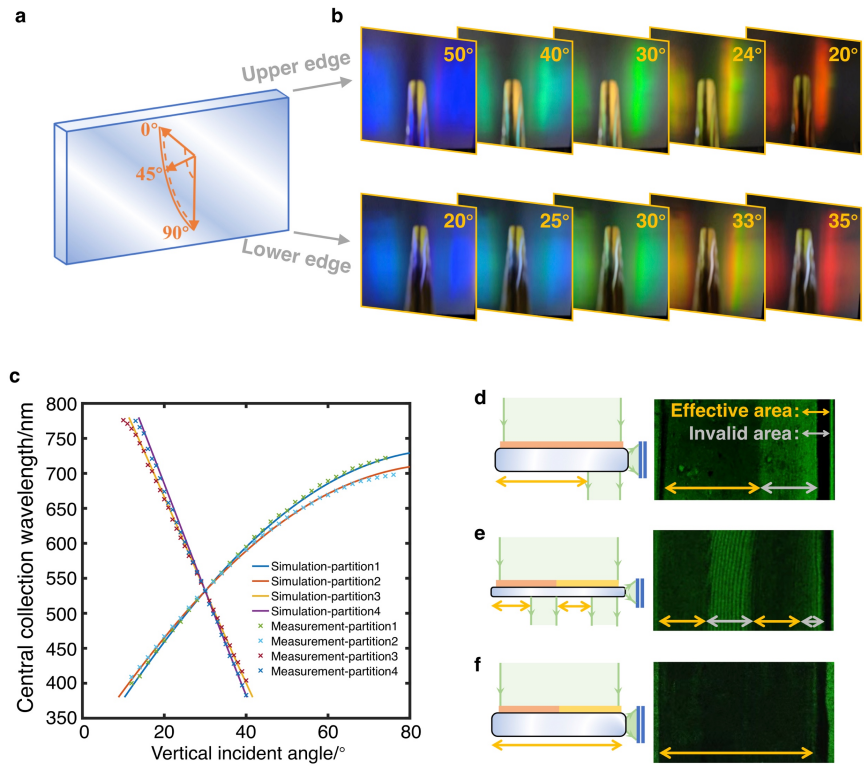

**Fig. 3 | Results of collecting characteristic analysis.**

**a**, Definition of vertical incident angle. The range of 0°-90° of vertical incidence angle is defined as the change from incidence perpendicular to the waveguide to incidence parallel to the waveguide. **b**, The collected light results on the upper and lower edges of the waveguide at different vertical incidence angles. Only 532nm green light is used for fabrication, while the different visible spectrum can be collected at different vertical angles. **c**, The relationship between the vertical incidence angle and the center collection wavelength. HOE-1 is divided into four partitions.

The specific parameters designed for different partitions vary, so the collection results of these four partitions are different. **d**, Secondary diffraction phenomenon on thick substrates without partitioning. **e**, Secondary diffraction phenomenon on thin substrates with partitioning. **f**, Elimination of secondary diffraction phenomenon on thick substrates with partitioning.

By utilizing the grating merging effect, the reconstructed conditions of HOE are not limited to the recording wavelength and angle any more. Figures 3a and 3b show the definition of the vertical incidence angle and the variation of the collected light at the upper and lower edges of the waveguide when white light is illuminated at different vertical incidence angles on the optical combiner. As shown in Figure 3b, the wavelength of the effective collected light is greatly expanded from 532nm to entire visible spectrum. Figure 3c shows the detailed simulation and measurement results of the vertical incident angle and central collection wavelength for four different partitions of HOE-1. The experimental results are consistent with the simulation. For the design wavelength of 532nm, the effective collection of incident angle width is only about 3°. After optimization, the angle of the effective collected light is expanded from 3° to 60°. Besides, thanks to the larger vertical selection angle of HOE, the optical combiner also has a larger collection angle of approximately 106° in the horizontal direction. Overall, HOE-1 can collect solar energy within a large angle range of 106°× 60°, and the collection spectral band covers the entire visible spectrum.

Generally speaking, if the thickness of the waveguide or the total reflection angle is not large enough, the secondary diffraction effect will lead to a decrease in the effective area of HOE-1. Figures 3d, e, and f illustrate the effective length of collection area under different conditions by showing the uncollected transmitted light caused by secondary diffraction, taking 532nm parallel light incident and total reflection transmission towards the right edge of the waveguide as an example. As shown in Figure 3d and e, if the HOE-1 is not partitioned or the thickness of the waveguide is thin, the effective length is about 17.5mm and 15.5mm respectively. By comparison, as shown in Figure 3f, with a thick waveguide and partitioned HOE-1, the secondary diffraction effect is eliminated, resulting in an increase in the effective length to 25mm. It should be noted that HOE-1 can collect solar energy to both upward and downward directions, so the overall effective length is 50mm consequently.

Finally, we conduct tests on the collection and power supply efficiency of HOE-1. The results indicate that the collection efficiency of HOE-1 exceed 60% at all wavelengths of 457nm, 532nm, and 639nm. It should be pointed out that the collection efficiency is the ratio of the collected light power to the incident light power, therefore it is influenced by both the diffraction efficiency of HOE and the transmission efficiency of total reflection. Additionally, solar cells with a maximum output voltage of 5V are utilized to assess the power supply efficiency under sunlight. The collected solar energy can increase the output voltage from 0.63V to 3.94V, and the total collected solar energy can charge 205mAh in half an hour. The proposed system can fulfill the function of self-charging using solar energy with a lightweight structure.

## 3 Discussion

This paper proposes a lightweight self-charging holographic near-eye display system that collects sustainable solar energy over a large angle range without adding complex structures. The self-charging capability avoids the heavy weight, continuous heating, and frequent charging of huge-capacity batteries. Furthermore, there will be many interesting studies in the future, given that the power supply efficiency of current prototype needs further improvement. On one hand, the optical multiplexing characteristics[26,27] and more flexible light modulation schemes of HOE are anticipated to increase the collected solar energy. On the other hand, the collection of wasted signal light that is not received by users will also significantly improve power supply efficiency. Finally, solar cells with high conversion efficiency and power-per-weight[28,29], such as perovskite cells[30,31], will further enhance the output power. It is worth mentioning that the proposed self-charging method is not limited to the above-mentioned system, but is universally applicable. We believe that our contribution will help sustainable energy become the primary energy source for lightweight near-eye display system, enabling all-day wear and facilitating its widespread application.

## 4 Methods

### 4.1 Details of HOE design

The optical characteristics of solar capture HOE are crucial for solar energy collection. Two important indicators are effective collection area and collection angle. On one hand, the size of the effective collection area is limited due to the phenomenon of secondary diffraction caused by the uniform structure of HOE. As a solution, we partition HOE-1 into four areas and design different total internal reflection angles for each area, avoiding the premature coupling-out of collected solar energy. On the other hand, the angle selectivity of the volume HOE greatly limits the collection angle of HOE-1. Therefore, we design the recording wavelength and angle reasonably by k-vector analysis , and extend the collection wavelength from the recording wavelength to the entire visible spectrum. Based on the variable wavelength reconstruction condition of HOE, the expansion from collection wavelength to collection angle is successfully achieved.

### 4.2 Fabrication of HOE

In the proposed system, HOEs are attached to the same waveguide to facilitate the coupling of light between air and waveguide. Therefore, to provide the lights with angles that meet the total reflection condition in waveguide, the fabrication inevitably requires prisms. For HOE-1, we utilize a right-angle prism to achieve exposure in two steps by employing three beams. Meanwhile, HOE-2 and HOE-3 are fabricated using an equilateral prism. Additionally, the angles of the exposure beams are precisely calculated based on the required angles in air, taking into account the effects of refraction and total reflection within the prism. It should be noted that HOE-1 undergoes substrate transfer after fabrication and is attached to the outer surface of the substrate where HOE-2 and HOE-3 are located.

**4.3 Implementation of prototype**

The experimental prototype can be divided into three parts: near-eye display module, solar energy collection module, and power supply module. Firstly, in the near-eye display module, a color laser projection with an integrated collimation system are implemented as the image source and the signal light emitted from it is perpendicular to the waveguide surface. Besides, the widths of the fabricated HOE-2 and HOE-3 are both 7mm. The thickness of the glass waveguide is 5mm. Secondly, in the power supply module, the area of the fabricated HOE-1 is equal to that of the waveguide. Two silicon-based solar cells with a maximum output voltage of 5V are located on the upper and lower sides of the waveguide respectively. Finally, the output electrical energy of the solar energy collection module is transmitted through wires as input energy for the near-eye display module.

## Acknowledgments

This work was financially supported by National Natural Science Foundation of China (Grant No. U22A2079 and 62035003); Beijing Municipal Science & Technology Commission, Administrative Commission of Zhongguancun Science Park (Grant No. Z211100004821012).